\documentclass[aps,prb,reprint,superscriptaddress,amsmath,amssymb,floatfix]{revtex4-2}
\usepackage[utf8]{inputenc}
\usepackage[T1]{fontenc}
\usepackage{graphicx}
\usepackage{dcolumn}
\usepackage{bm}
\usepackage{hyperref}

\begin{document}

\title{Interactions between structural and magnetic domains through the Verwey transition in stoichiometric and doped magnetite}

\author{Mateusz \'{S}l\k{e}zak}
\author{Zbigniew K\k{a}kol}
\author{Ryszard Zalecki}
\author{Andrzej Koz\l{}owski}
\affiliation{AGH University of Krakow, Faculty of Physics and Applied Computer Science, al. Mickiewicza 30, 30-059 Krakow, Poland}

\author{Naveen Kumar Chogondahalli Muniraju}
\affiliation{The Henryk Niewodniczanski Institute of Nuclear Physics, Polish Academy of Sciences, ul. Radzikowskiego 152, 31-342 Krakow, Poland}

\author{Neven Bari\v{s}i\'{c}}
\affiliation{Institute of Solid State Physics, TU Wien, 1040 Vienna, Austria}
\affiliation{Department of Physics, Faculty of Science, University of Zagreb, Bijenicka cesta 32, HR-10000 Zagreb, Croatia}

\author{Stuart Gilder}
\affiliation{Department of Earth and Environmental Sciences, Ludwig Maximilian University, Munich, Germany}

\author{Wojciech Tabi\'{s}}
\email{wtabis@agh.edu.pl}
\affiliation{AGH University of Krakow, Faculty of Physics and Applied Computer Science, al. Mickiewicza 30, 30-059 Krakow, Poland}

\begin{abstract}
The magnetic response of magnetite as it cycles through the Verwey transition at $T_{\rm V}\approx 120$ K remains incompletely understood, despite numerous studies of both structural and magnetic domains. For example, although the alternating-current susceptibility increases upon heating through the Verwey temperature $T_{\rm V}$, the magnetization may either increase or decrease. Since many intertwined processes occur at $T_{\rm V}$, a multifaceted view of the magnetic changes may help clarify the nature of these processes and establish specific features of the transition as diagnostic tools. To this end, we report measurements of the magnetic response of magnetite upon heating through $T_{\rm V}$ under a wide range of applied-field conditions. We studied synthetic stoichiometric and doped (Zn, Ti, and Al) single crystals, as well as a natural polycrystalline sample, with the aim of explaining how and why the magnetization changes as a function of temperature through $T_{\rm V}$. The magnetic-field-induced reorientation of the monoclinic $c$ axis may broaden the transition, primarily in stoichiometric magnetite. These findings help clarify the mechanism of the transition.
\end{abstract}

\maketitle

\section{Introduction}

Magnetite Fe$_3$O$_4$ helps guide organisms either through an external guidance system, such as a man-made compass, or through internally driven magnetoreception, as in magnetotactic bacteria \cite{baumgartner2013}. Biogenic magnetite nanoparticles have great potential in medicine and pharmacology as drug-delivery systems \cite{holban2014}, and magnetite-bearing chiton teeth provide nearly single-domain-sized magnetite for research applications \cite{seifert2017}. Upon cooling from the paramagnetic state through the Curie temperature, $T_{\rm C}\simeq 853$ K, magnetite can acquire a remanent magnetization related to the geomagnetic-field direction, thereby enabling paleomagnetists to study geological processes and the geodynamo over billions of years of Earth's history. Magnetite also serves as a magnetic recorder in memory-storage devices. Owing to its spin-polarized transport, magnetite is a promising candidate for spin electronics \cite{gupta1999,schmitt2021}; de Jong \textit{et al.} \cite{de2013} demonstrated that laser excitation can switch magnetite across the insulator--metal transition at $T_{\rm V}$ from its low-temperature insulating state to its high-temperature metallic state within femtoseconds, making magnetite a candidate material for ultrafast switches. Recent studies of thin films grown on different substrates \cite{feo2016,enhanced2025} have shown that $T_{\rm V}$ can be controlled, either lowered or raised, thereby broadening the range of possible spintronic applications.

This metal--insulator phase transformation is known as the Verwey transition (VT) \cite{verwey1939}, although related low-temperature anomalies had already been reported earlier by Karl Renger in work associated with Pierre Weiss and Albert Einstein \cite{renger1913}. The transition results from the interplay of electronic, lattice, and magnetic subsystems. As recently suggested, magnetic interactions shape the properties of magnetite and strongly affect the VT \cite{podgorska2025}. Thus, the Verwey transition is a sensitive probe of magnetite properties not only near the Verwey temperature $T_{\rm V}$, but also at room temperature and even higher temperatures.

Above the VT, magnetite has cubic $Fd\bar{3}m$ crystal symmetry, with two magnetic sublattices: tetrahedral A sites occupied solely by Fe$^{3+}$ atoms and octahedral B sites with mixed-valence Fe$^{3+}$ and Fe$^{2+}$ atoms. The magnetic moments of these two sublattices are antiparallel. The easy magnetic axes lie along $\langle 111\rangle$ for $T>130$ K, with the hard and intermediate axes along $\langle 100\rangle$ and $\langle 110\rangle$, respectively, as shown in Fig.~S1 in the Supplemental Material (SM). Below the isotropy point $T_{\rm IP}$, where the main anisotropy constant $K_1$ is zero, the easy axes lie along the cubic edges down to $T_{\rm V}$, where the anisotropy energy increases by a factor of about 100 \cite{abe1976,koenitzer1992}. In nonstoichiometric, oxidized magnetite, Fe$_{3(1-\delta)}$O$_4$, or in doped magnetite such as Fe$_{3-x}$Zn$_x$O$_4$ (Fig.~S1), this increase in anisotropy energy is lower.

Below the VT, magnetite has monoclinic $Cc$ symmetry \cite{senn2012}, with the magnetic easy axis lying parallel to the monoclinic $c$ axis, which is oriented along one of the high-temperature cubic $\langle 100\rangle$ edges (Fig.~S1a). In total, 24 structural domains may form below $T_{\rm V}$ \cite{kasama2010}, unless a magnetic field is applied along one of the $\langle 100\rangle$ axes during cooling, which reduces the number of possible domains. Interestingly, the cubic $Fd\bar{3}m$ crystal symmetry above $T_{\rm V}$ represents only an average structure, whereas locally, within a single unit cell, the symmetry resembles that of the low-temperature structure \cite{perversi2019,piekarz2021}.

The many types of structural domains at $T<T_{\rm V}$, with different monoclinic $c$ axes, together with the variety of magnetic domains linked to them \cite{kasama2010,de2019,martngarca2018,kubo2015,nagy2019}, affect the magnetization depending on whether a magnetic field is applied and, if so, on its strength and direction, as seen in measurements of magnetic moment $\mu$ versus magnetic field $B$. In particular, when magnetite is warmed through $T_{\rm V}$ in a magnetic field, the $c$ axis may switch below $T_{\rm V}$ toward the cubic $\langle 100\rangle$ direction closest to the magnetic-field direction. This process is known as axis switching (AS) \cite{calhoun1954,kolodziej2023} and is illustrated in Figs.~S2a--S2c. The effect of AS was recently mentioned in a study of a 200-nm Fe$_3$O$_4$ film grown on a Mg$_2$TiO$_4$ substrate \cite{enhanced2025}; however, that work considered only the change of the magnetic easy axis, without taking into account the accompanying structural changes. We have previously studied axis switching using macroscopic methods, including VSM \cite{krol2007} and measurements under hydrostatic pressure \cite{kolodziej2023}, as well as microscopic techniques such as NMR \cite{chlan2010}, XRD, and M{\"o}ssbauer spectroscopy \cite{kolodziej2020}. Among the most important parameters describing AS are the activation energy $U$ and the switching field $B_{\rm SW}$, the latter being strongly dependent on temperature and stoichiometry. Since $B_{\rm SW}$ plays a key role in understanding the results presented below, the temperature dependence of $B_{\rm SW}$ for different doping levels is recalled in Fig.~S3 \cite{kolodziej2023}.

This article addresses the impact of axis switching on the magnetization process through the VT. We show that, in stoichiometric magnetite, the main influence of AS is that the magnetic moment $\mu$ measured upon heating in fields on the order of 0.1 T changes already below the true $T_{\rm V}$. This makes the VT appear broadened, as would be expected for an inhomogeneous sample. We also study features of $\mu$ vs. $T$ that were previously observed in pressure-cycling experiments on natural, synthetic, and doped magnetite \cite{bialo2019}. In those measurements, the magnetization was found either to increase or to decrease upon warming through the Verwey transition, presumably depending on internal strain and defects. Here, we examine the change in $\mu$ with temperature, i.e., the $\mu(T)$ step, upon heating samples from $T<T_{\rm V}$. We find that $\mu$ decreases upon heating, producing a down-step, when the sample is measured in zero or very small fields ($<1$ mT). By contrast, $\mu$ increases upon heating, producing an up-step, when the measuring field is 10 mT or 0.18--0.25 T, similar to the behavior observed in dynamic susceptibility measurements $\chi_{\rm AC}$, where the ac magnetic field is about 1 mT. Possible explanations of these effects are provided below.

\section{Experiments}

We studied seven samples: synthetic single crystals ranging from stoichiometric ($\delta=0$) to slightly oxidized ($\delta=0.0025$) magnetite, as well as crystals doped with various amounts of Ti, Al, and Zn, all grown by the skull-melter method \cite{harrison1978,aragon1983,aragon1982,wang1987}, together with one natural polycrystalline sample provided by the Bavarian State Mineralogy Museum, Munich, Germany (See Table~\ref{tab:samples} and S4 in SM for details). The Verwey temperatures $T_{\rm V}$ were determined from dynamic susceptibility measurements (Fig.~S4a), and the transition widths were estimated from the pronounced changes in $\chi_{\rm AC}(T)$. The sample compositions were determined mainly from the $T_{\rm V}$ vs. $x$ or $3\delta$ relation shown in Fig.~S4b \cite{kozlowski1996,kakol2015}; the composition of the natural magnetite sample was determined by x-ray microprobe analysis \cite{bialo2019}.

\begin{table*}[t]
\caption{\label{tab:samples}Composition $x$ or nonstoichiometry parameter $\delta$, Verwey temperature $T_{\rm V}$, transition width $\Delta T_{\rm V}$ determined from ac susceptibility $\chi_{\rm AC}$, and mass $M$ of the measured samples. Magnesium is the main dopant in the natural polycrystalline sample.}
\begin{ruledtabular}
\begin{tabular}{lllll}
Sample & $x$, $\delta$ & $T_{\rm V}$ ($\chi_{\rm AC}$) (K) & $\Delta T_{\rm V}$ ($\chi_{\rm AC}$) (K) & $M$ (mg) \\
\hline
Ti-doped, 447.4.2 & 0.012 & 102.1 & 4.3 &  \\
Ti-doped, 447\#2A & 0.010 & 111.2 & 2.4 & 7.5 \\
Nonstoichiometric, $\delta=0.0025$ & 0.0025 & 113.8 & 1.4 & 156.5 \\
Al-doped, 432-2 & 0.014 & 112.9 & 4.1 & 24.9 \\
Stoichiometric, mg05N\#2 & 0 & 123.1 & 0.3 & 17.0 \\
Natural, 17909 & 0.002 (Mg) & 120.5 & 6.4 & 23.5 \\
Zn-doped, 448\#1BW & 0.0067 & 114.7 & 1.2 & 15.6 \\
\end{tabular}
\end{ruledtabular}
\end{table*}

Measurements of $\mu(T)$ were performed with a Lake Shore 7300 VSM in a magnetic field of 0.9 T. The samples were cylindrical, with diameters from 0.9 to 3 mm and the cylinder axis along [001] (Fig.~S2d). Throughout the manuscript, we adopt the high-temperature cubic notation both above and below $T_{\rm V}$. Thus, the monoclinic $c$ axis refers to the [100] axis, along which the magnetic field was applied during cooling through $T_{\rm V}$. The relative uncertainty of the data points for each sample is below 1\%, approximately the symbol size. The mean absolute uncertainty is about 8\%.

\begin{figure*}[t]
\centering
\includegraphics[width=0.99\textwidth]{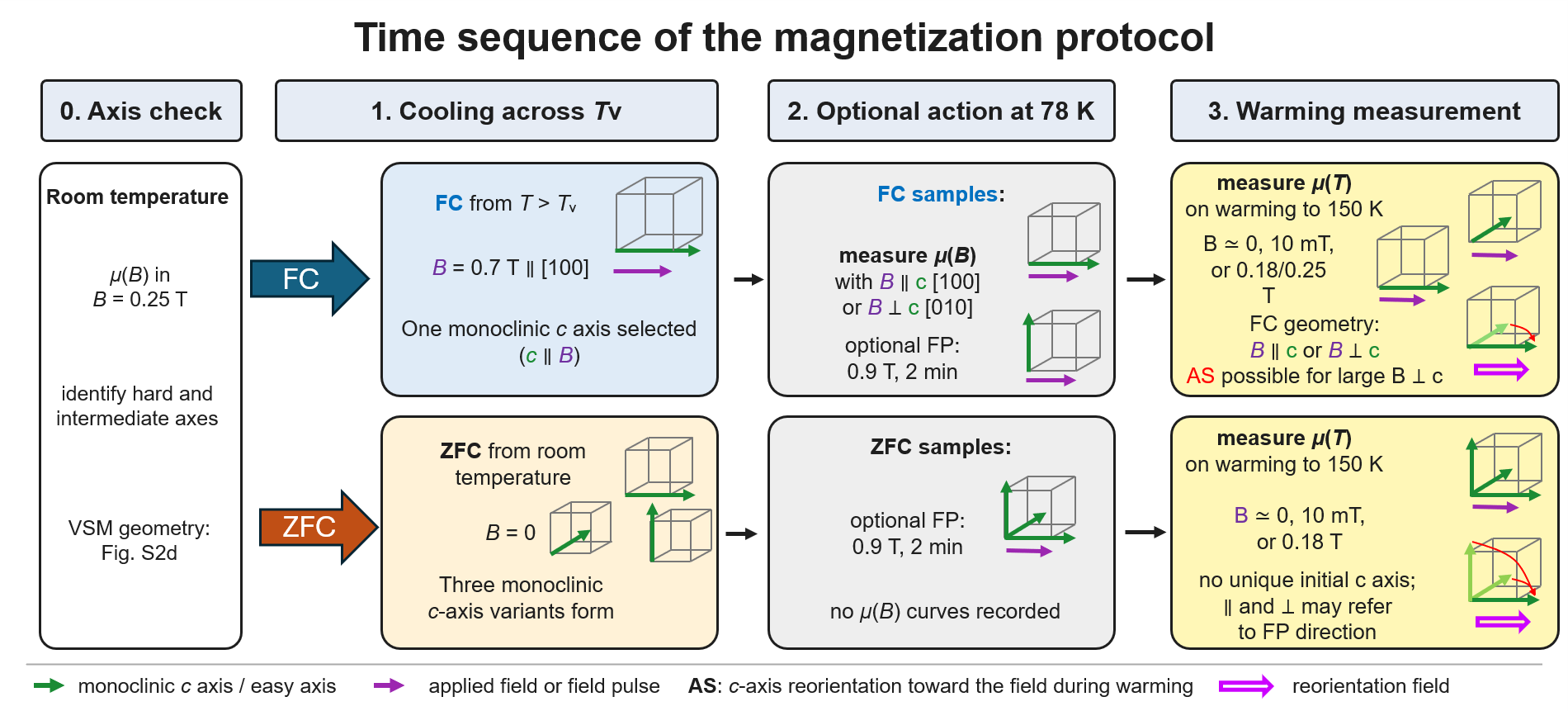}
\caption{\label{fig:experiment}Summary of the experimental protocols used for the magnetization measurements. The sample was first cooled from above the Verwey transition either in a magnetic field (FC, upper panels), or in zero field (ZFC, lower panels), down to 78 K. In the FC protocol, a field of 0.7 T was applied along the cubic [100] direction, thereby selecting a single monoclinic $c$ axis below $T_{\rm V}$, which is also the magnetic easy axis. This selected $c$ axis is indicated schematically by the dark-green arrows in the cubes. The purple arrows indicate the direction of the applied magnetic field: thick double purple arrows denote a larger field or the field pulse, whereas thin single purple arrows denote a smaller applied field. In the ZFC protocol, no field was applied during cooling, so several monoclinic $c$ axes and corresponding structural domains could form, as schematically shown in the lower panel. After cooling, an additional action could be performed at $T<T_{\rm V}$, as indicated in the gray panels. For FC samples, the magnetic moment $\mu(B)$ was measured either with $B\parallel c$ or with $B\perp c$. In some cases, a magnetic field pulse FP, typically 0.9 T for 2 min, was also applied at 78 K. For ZFC samples, an FP was applied, but without recording $\mu(B)$ curves. The notation used throughout the manuscript follows this sequence and color code. For example, FC, 80 K, $B\perp$ denotes a sample field-cooled to 78 K and then measured with the magnetic field perpendicular to the $c$ axis imposed during FC.}
\end{figure*}

The experimental procedure is summarized in Fig.~\ref{fig:experiment}, and a more detailed description of the protocol sequence along with the notation is given in Sec.~S5 and Fig.~S5 of the SM. Each sample was first measured at room temperature in an external magnetic field $B=0.25$ T to identify the hard and intermediate magnetic axes; the VSM geometry is shown in Fig.~S2d. After the orientations were determined, the sample was either FC to 78 K in $B=0.7$ T applied along [100], thereby forcing the monoclinic $c$ axis and the magnetic easy axis to be parallel to $B$, or ZFC from room temperature to 78 K. These two cooling protocols are shown in the upper and lower panels of Fig.~\ref{fig:experiment}, respectively.

The subsequent action at $T<T_{\rm V}$ depended on the sample. Our experience is that even small amounts of dopants, as in the present study, drastically increase the switching field $B_{\rm SW}$ needed to reorient the $c$ axis, as shown in Fig.~S3 \cite{kolodziej2023}. Therefore, for doped samples, $\mu(B)$ was measured first along the direction perpendicular to the $c$ axis, [010], and then along the $c$ axis, [100]. For the purely stoichiometric sample, the structural-domain state is sufficiently ``weak'' that $\mu(B)$ along [010] was not measured, because AS occurs immediately. Additionally, the samples at 78 K could be subjected to a 0.9 T field for 2 min, referred to as a field pulse (FP) in column 2 of Fig.~\ref{fig:experiment}, to reproduce the procedure of Ref.~\cite{kolodziej2020}.

After these steps at $T<T_{\rm V}$, the magnetic moment $\mu$ vs. $T$ was measured up to 150 K in various fields applied either along or perpendicular to the easy axis. This constituted the main part of our measurements and is shown in column 3 of Fig.~\ref{fig:experiment}. This part of the figure, together with the extended presentation in Sec.~S5 and Fig.~S5 of the SM, schematically defines the measurement procedure and illustrates possible $c$-axis switching. For ZFC samples, shown in the lower panel of Fig.~\ref{fig:experiment}, the heating directions are not specified, although the symbols $\parallel$ and $\perp$ may refer to the field-pulse direction.

\section{Results}

\subsection{High-field response and axis switching}

\begin{figure*}[t]
\centering
\includegraphics[width=\textwidth]{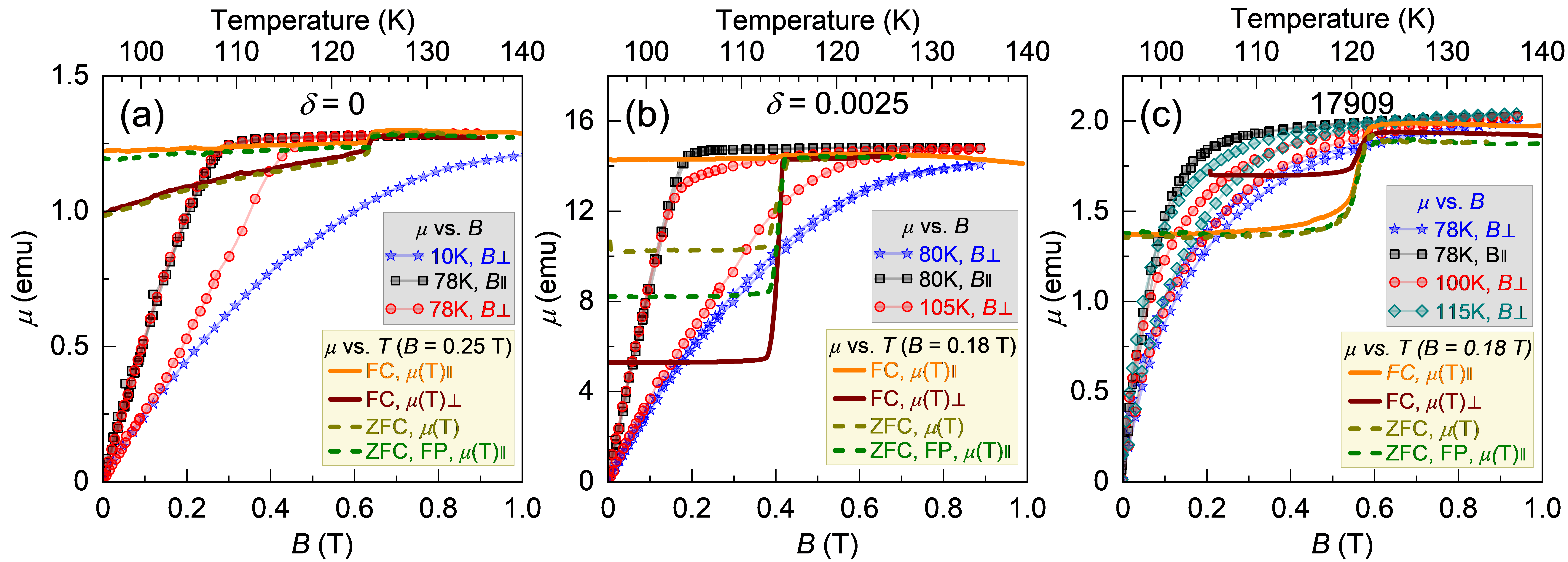}
\caption{\label{fig:main2}Isothermal $\mu(B)$ and in-field $\mu(T)$ across the Verwey transition for stoichiometric Fe$_3$O$_4$ ($\delta=0$, a), nonstoichiometric Fe$_{3(1-\delta)}$O$_4$ ($\delta=0.0025$, b), and natural polycrystalline magnetite (17909, c). Each panel combines $\mu(B)$ isotherms (symbols, bottom axis; legend on gray background) with $\mu(T)$ warming curves (lines, top axis; legend on yellow background) measured in $B=0.25$ T (a) or 0.18 T (b, c) after the cooling protocols of Fig.~\ref{fig:experiment}. Throughout, $\parallel$ and $\perp$ denote the orientation of the measuring field relative to the field applied during field cooling or the field pulse; $B\parallel$ and $B\perp$ thus correspond to sweeps along [100] and [010], respectively.}
\end{figure*}

The initial step in each $\mu$ vs. $T$ experiment presented below was to measure $\mu(B)$ at several temperatures, starting at 78 K. When the monoclinic $c$ axis had been aligned during FC, the first measurements were performed along this axis, [100]. To check for possible AS, $\mu(B)$ was then measured perpendicular to it, along [010]. The results for the stoichiometric sample mg05N\#2, the nonstoichiometric sample with $\delta=0.0025$, and the natural polycrystalline magnetite sample are shown in Fig.~\ref{fig:main2}. In this figure, the $\mu(T)$ curves measured in 0.18--0.25 T are superimposed on the corresponding $\mu(B)$ results. Additional $\mu(B)$ data for the remaining samples are shown in Fig.~S6 of the SM, while the corresponding superpositions of $\mu(B)$ and $\mu(T)$ are shown in Figs.~S7 and S8.

For the stoichiometric sample, the irreversible jump in $\mu(B)$ along [010], which signals AS, is already observed at 78 K. In contrast, for the nonstoichiometric and natural samples, it occurs only much closer to $T_{\rm V}$, for example at 105 K for $\delta=0.0025$. The overlay in Fig.~\ref{fig:main2} shows directly how AS, occurring at a specific combination of field and temperature, shapes $\mu(T)$. For $\delta=0$, the FC, $\mu(T)_\perp$ curve increases continuously from below 78 K due to AS, whereas for the nonideal samples the AS contribution is confined to approximately 10 K below $T_{\rm V}$. In all cases, $\mu(T)$ measured in these fields ends with an upward step, or up-step, at $T_{\rm V}$.

\subsection{Low-field reversal of the magnetic step}

\begin{figure*}[t]
\centering
\includegraphics[width=\textwidth]{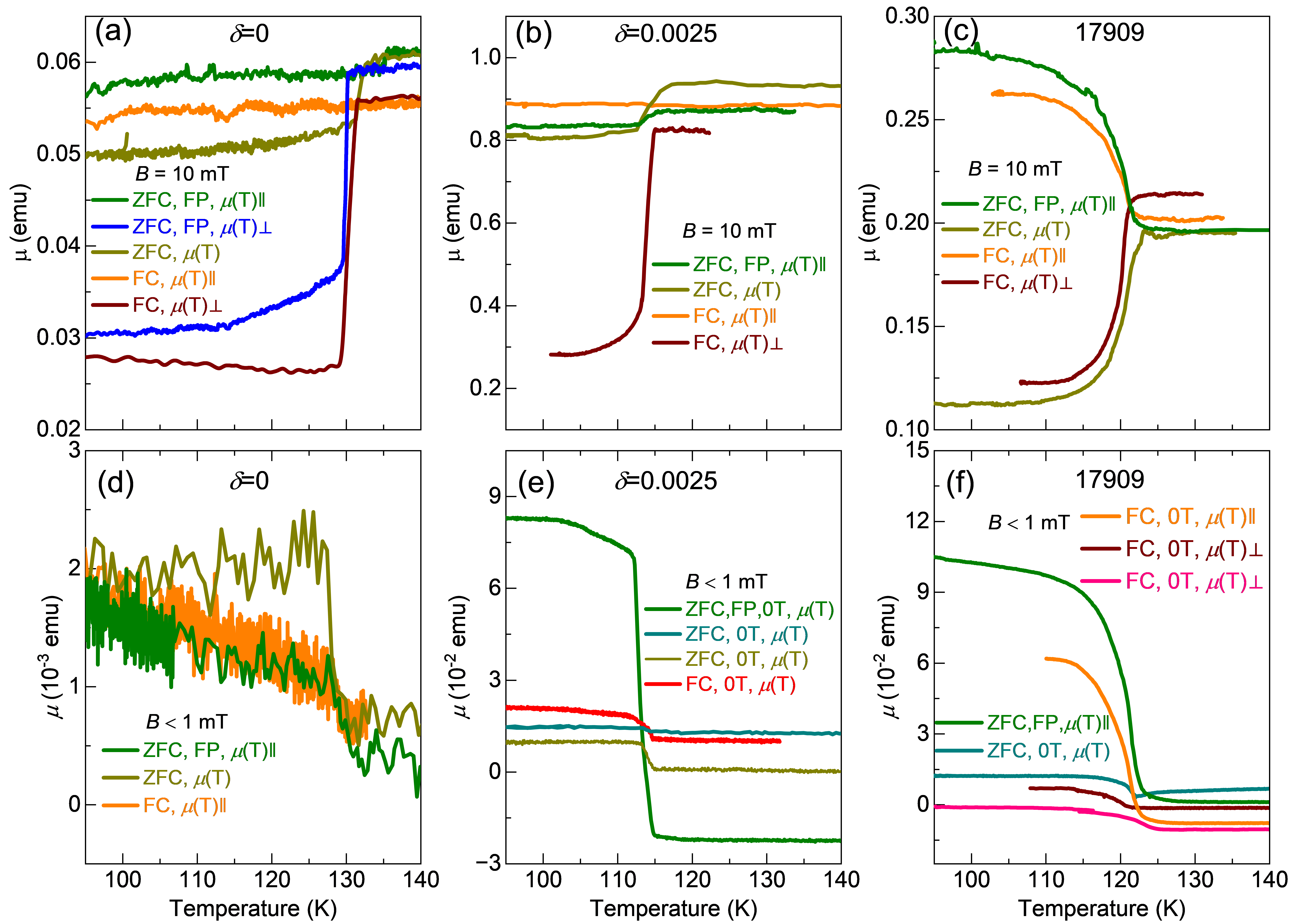}
\caption{\label{fig:main3}Low-field $\mu(T)$ on warming through $T_{\rm V}$. Columns: $\delta=0$ (a, d), $\delta=0.0025$ (b, e), and 17909 (c, f); color code and protocol labels follow Fig.~\ref{fig:experiment}. Top row (a--c): $B=10$ mT. Bottom row (d--f): near-zero field, $B<1$ mT; curves labeled 0T were measured in nominally zero field and represent the remanent moment.}
\end{figure*}

The corresponding low-field results are presented in Fig.~\ref{fig:main3}. In $B=10$ mT (Figs.~\ref{fig:main3}a--c), the step at $T_{\rm V}$ is generally upward, as in higher fields, although for the natural sample measured after FC with $B$ along the $c$ axis, a down-step appears instead (Fig.~\ref{fig:main3}c). In near-zero field, $B<1$ mT (Figs.~\ref{fig:main3}d--f), the sign of the step reverses: the magnetic moment decreases upon heating for all samples and cooling protocols. This field-dependent crossover of the step at $T_{\rm V}$, from a down-step at $B\lesssim 1$ mT to an up-step for $B\geq 20$ mT, is one of the key results of this study and is observed consistently across stoichiometric, nonstoichiometric, and natural samples (see also Figs.~S7 and S8).

When comparing the results, it is important to remember that the formation conditions for the natural sample 17909 and the synthetic samples are drastically different. Natural magnetite grows over geological time scales, whereas synthetic crystals form within a day. This difference is most evident in their morphology: natural single crystals are octahedral, with well-defined crystal-lattice directions, whereas synthetic crystals have no defined morphology (Fig.~S9). Despite this underlying difference, our results do not differentiate between natural and synthetic samples to the extent suggested by the morphology, although some differences can be seen.

The main aim of this work was to determine how magnetization measurements through the Verwey transition are affected by experimental conditions, in particular by the strength and direction of the applied magnetic field and by the cooling protocol. The two most striking features of the $\mu(B)$ and $\mu(T)$ results are the reversal of the step direction at $T_{\rm V}$, from a down-step at $B\simeq 0$ to an up-step for $B\geq 10$ mT (Fig.~\ref{fig:main3}), and the occurrence of AS (Fig.~\ref{fig:main2}). These effects, together with other subtleties of $\mu(B)$ and $\mu(T)$, are discussed below.

\section{Discussion}

\subsection{Axis switching and apparent broadening of the transition}

As noted above, the $\mu(T)$ characteristics of magnetite depend strongly on how the sample is cooled below $T_{\rm V}$ and on the field applied during $\mu$ vs. $T$ measurements. This dependence is most clearly reflected in the role of axis switching and in the sign and shape of the $\mu$ vs. $T$ step at $T_{\rm V}$.

The main impact of axis switching on $\mu(T)$ is observed in the stoichiometric single crystal (Fig.~\ref{fig:main2}a). In a magnetic field of 0.25 T applied along [010], i.e., along a non-$c$ direction after the monoclinic $c$ axis had been established during FC, the $\mu$ vs. $T$ signal does not show an abrupt step. Instead, it changes continuously already from the lowest measured temperature, 78 K. As a result, the apparent $T_{\rm V}$ region in stoichiometric magnetite appears broadened, which could be mistakenly interpreted as arising from an inhomogeneous distribution of dopants or nonstoichiometric regions.

\begin{figure}[t]
\centering
\includegraphics[width=0.5\textwidth]{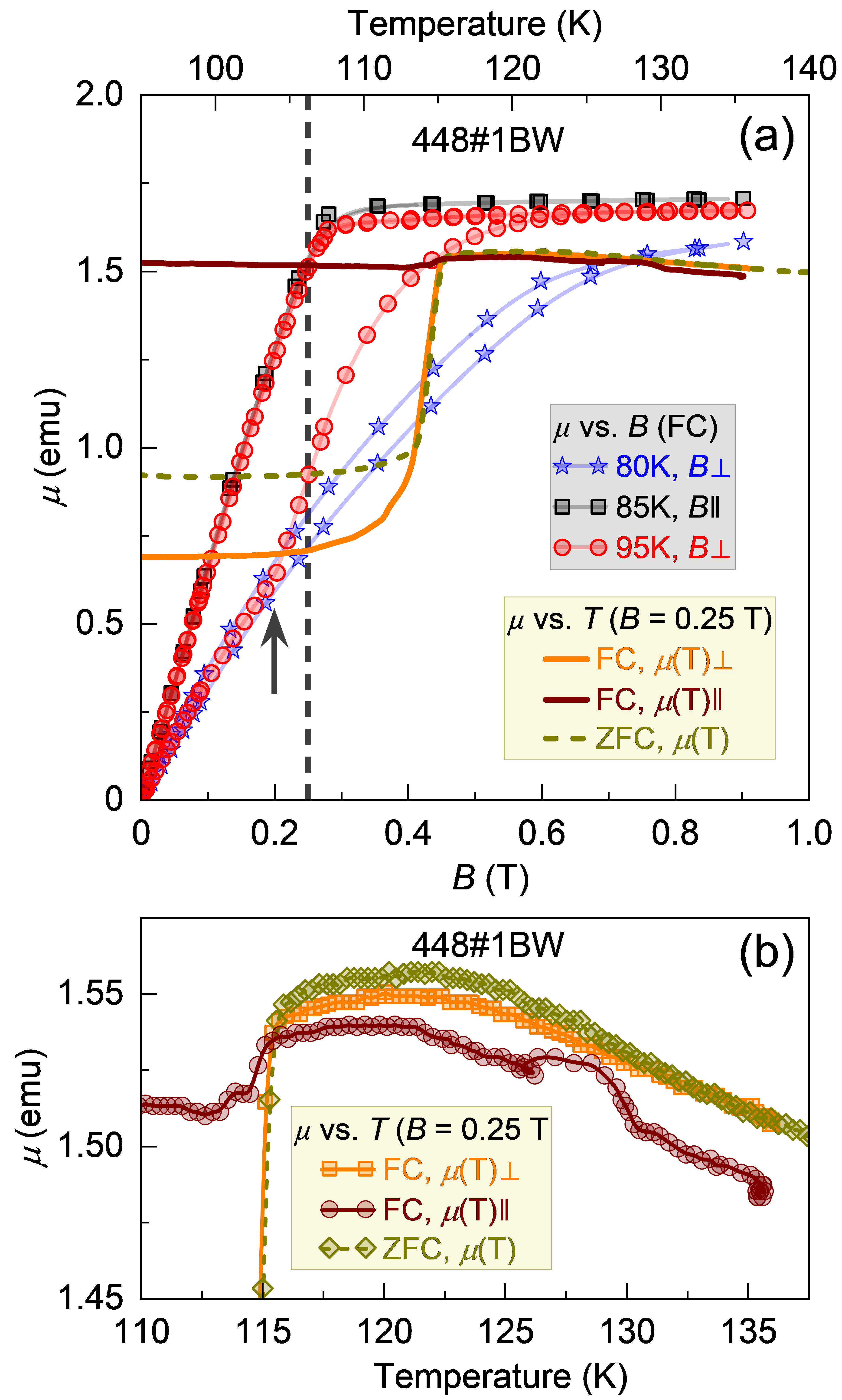}
\caption{\label{fig:main4}Magnetization of the Zn-doped magnetite sample (448\#1BW). (a) $\mu(T)$ measured in $B=0.25$ T in the vicinity of $T_{\rm V}$ after the cooling protocols of Fig.~\ref{fig:experiment}, superimposed on $\mu(B)$ isotherms; the arrow marks AS at 95 K near $B=0.2$ T, and the vertical dashed line marks the 0.25 T measuring field. (b) Fine structure of $\mu(T)$ at 0.25 T from just below to approximately 30 K above $T_{\rm V}$.}
\end{figure}

The influence of the $c$-axis orientation on $\mu(T)$ in fields of 0.18 and 0.25 T for doped and nonstoichiometric samples is illustrated in Fig.~\ref{fig:main4} for the Zn-doped sample 448\#1BW.

For $B$ applied along [010], i.e., perpendicular to the $c$ axis selected during FC along [100], the initial signal is lower because the magnetic moment tends to align with the easy axis [100] rather than with the applied field. This corresponds to the orange curve in Fig.~\ref{fig:main4}a, labeled FC, --, 0.25 T, $\mu(T)_\perp$, where ``--'' denotes no additional field pulse or low-temperature field treatment before warming, and $\perp$ indicates that the measuring field is perpendicular to the $c$ axis imposed during FC, following the notation of Fig.~S5. However, when $T$ approaches $T_{\rm V}$ from below, at about 106 K, axis switching begins and the [010] direction starts to become an easy axis. This is visible as a slight gradual increase in the signal, followed at higher temperature by the same changes as in the red curve. Thus, the configuration in which $B$ is applied along [010] during heating after FC along [100] produces the largest step at $T_{\rm V}$.

Finally, after ZFC, when only one third of the $c$-domain variants are aligned with the field direction, the initial signal (dashed olive curve in Fig.~\ref{fig:main4}a; ZFC, --, 0.25 T) lies between the orange and brown curves, and the step at $T_{\rm V}$ reaches approximately $2/3$ of its maximum value. As $T$ increases, its behavior resembles that of the red curve, as expected.

In lower fields, changes of the $c$ axis are small; nevertheless, possible AS should be taken into account when interpreting subtle features of $\mu$ vs. $B$. Figure~\ref{fig:main4}b shows the fine structure of $\mu(T)$ at 0.25 T for a representative Zn-doped sample. The very slow rise up to 128 K is associated with the decrease of the main anisotropy constant, $K_1$, toward the isotropy point $T_{\rm IP}$, above which the easy magnetization axes change from $\langle 100\rangle$ to $\langle 111\rangle$. At higher temperatures, the signal decreases steadily, consistent with the increase of the anisotropy constant.

The net outcome of our results is that the true $T_{\rm V}$ should not be defined as the mean temperature derived from the lowest and highest values of the $\mu$ signal, because these values are influenced by AS below $T_{\rm V}$ (Fig.~\ref{fig:main4}a) and by the evolution of anisotropy around $T_{\rm IP}$ above $T_{\rm V}$ (Fig.~\ref{fig:main4}b).

\subsection{Origin of the field-dependent sign of the magnetic step}

The step in $\mu$ vs. $T$ at $T_{\rm V}$ provides the clearest view of the $\mu(T)$ characteristics. In our experience, when $B$ is greater than 20 mT, this step always increases upon heating, whereas when $B$ is close to zero, below 1 mT, it always decreases (Figs.~\ref{fig:main3}d--f). In a slightly higher field of 10 mT, the step is usually upward, as in higher fields (Figs.~\ref{fig:main3}a--c). However, depending on the sample, it can also be downward either after the field pulse or when the measuring field is applied along [100], the $c$ axis selected during FC (Fig.~\ref{fig:main3}c).

Once the sample is cooled below $T_{\rm V}$, a magnetic domain structure is established under the combined influence of the new crystallographic structure, as the high-temperature cubic phase gives way to the monoclinic $Cc$ phase, the approximately 100-fold increase in anisotropy energy, and the cooling procedure. Although the $c$ axis is selected by the magnetic field applied during FC, so that the major structural $c$-domain variants are essentially absent, $a$-$b$ and $c$-$c$ structural domains remain present. The resulting structural-domain pattern is complicated, and these domains are small. The same applies to the magnetic domains, which are comparable to the structural ones. Both types of domains are pinned by defects arising from nonstoichiometry, as in the sample with $\delta=0.0025$, from doping, or, in the case of natural magnetite, from the different grains in this polycrystalline sample. Whatever the magnetic-domain structure is, its modification in fields $B>20$ mT remains difficult because of the large anisotropy and the abundance of structural domain walls and defects, all of which act as pinning centers. Above $T_{\rm V}$, however, the anisotropy decreases and the number of pinning centers is reduced, because structural domain walls are no longer present. As a result, more magnetic moments align with the magnetic field, the total magnetic moment increases, and a stepwise increase is observed.

The situation in the lowest applied fields, below 1 mT, is slightly different. After FC, the $c$ axis is essentially aligned with the field direction, and the material becomes magnetized to some extent. A similar situation occurs after a field pulse, which may force the $c$ axis along the field-pulse direction.

Since the anisotropy at $T<T_{\rm V}$ is rather strong, a small field of about 1 mT is insufficient to reverse the magnetic moments in the magnetized material. This effect is reinforced by defects that pin magnetic domain walls; both anisotropy and pinning act to anchor the magnetic moments because the applied field is too weak. Once $T_{\rm V}$ is reached and the anisotropy decreases, a magnetic field of 1 mT tends to align the magnetic moments along the field, as in the higher-field case, $B>20$ mT, described above. However, a 1 mT field is too weak for this effect to dominate, so the magnetic response is governed instead by dipolar interactions; eventually, the magnetic moments within domains become smaller. Consequently, in this small field, the total magnetic moment of the sample decreases. A related sensitivity of the magnetic signal at $T_{\rm V}$ was reported by Dunlop and {\"O}zdemir \cite{dunlop2018}, who studied remanence cycling of 0.6--135 $\mu$m magnetite particles across the Verwey transition. They observed that the remanence may either decrease or increase across $T_{\rm V}$, depending on particle size and on whether the saturation remanence was imparted at 10 K or 300 K. They attributed this behavior to the combined influence of anisotropy, domain structure, and crystal defects below and above the Verwey transition. This supports our conclusion that the sign of the magnetic step at $T_{\rm V}$ is not a simple intrinsic marker of the electronic transition alone, but results from the interplay between magnetic anisotropy, domain configuration, and pinning centers of structural or defect origin.

\subsection{Relation to ac susceptibility and magnetically defined $T_{\rm MV}$}

These different steps at $T_{\rm V}$ contrast with the dynamic susceptibility $\chi_{\rm AC}$, which always shows an up-step at $T_{\rm V}$ upon heating, even when the ac-field frequency is 0.2 Hz (see Sec.~S9 and Fig.~S10 of the SM for further discussion and $\chi_{\rm AC}$ results). The step in the magnetite $\chi_{\rm AC}$ data was discussed in terms of the competition between pinning centers and the magnetic field \cite{balanda2005}. Specifically, applying a small ac magnetic field results in magnetic-domain-wall movement (DWM). Since magnetic domains interact with crystallographic domains below $T_{\rm V}$, DWM is hindered, resulting in a lower $\chi_{\rm AC}$ signal at $T<T_{\rm V}$. At higher temperatures, crystallographic domains are absent, so DWM increases, and the signal increases accordingly \cite{balanda2005}.

Although the stepwise increase in $\chi_{\rm AC}$ is well understood, subtle effects of defects also influence $\chi_{\rm AC}$. This can be seen by comparing $\chi_{\rm AC}$ below $T_{\rm V}$ in stoichiometric magnetite \cite{balanda2005} with $\chi_{\rm AC}$ in doped and nonstoichiometric samples, as observed in more recent measurements \cite{tabis} and shown in Fig.~S10c. Notably, when measurements are performed after FC in $B=0.2$ T, a strong peak appears approximately 15 K below the transition temperature in the nonstoichiometric sample, which is not visible in stoichiometric magnetite or after ZFC. To the best of our knowledge, the presence of such a peak has not been reported, and it may be a simple indication of a defect field. In any case, the behavior of magnetic domains differs between ac and dc fields of identical strength, approximately 1 mT.

In Ref.~\cite{bialo2019}, it was suggested that the step in magnetic moment occurring at the transition temperature is distinct from the electronically driven Verwey transition at $T_{\rm V}$. This magnetically defined Verwey-transition temperature is referred to as $T_{\rm MV}$. It is a frequently observed parameter in geophysical studies of the Earth and marks the temperature at which the domain structure of the samples begins to change upon heating. The data presented above show that magnetic effects in magnetite, such as AS, DWM, and magnetic-domain-wall pinning centers, all of which are closely linked to the electronic Verwey transition, shape the transition when it is observed through magnetic phenomena and under magnetic fields of different strength. In extreme cases, either a strong magnetic field or very strong wall-pinning centers that prevent DWM may make $T_{\rm MV}$ invisible, which does not mean that the transition itself does not occur.

\section{Summary}

We performed extensive measurements of the temperature dependence of the magnetic moment, $\mu(T)$, upon warming magnetite through $T_{\rm V}$ in nominally zero field, 10 mT, and 0.18 or 0.25 T. The measuring field was applied either parallel or perpendicular to the easy axis selected during cooling. Before the warming measurements, the samples were prepared by FC, ZFC, or ZFC followed by a 0.9 T field pulse at 78 K along a $\langle 100\rangle$ direction.

The central observation of this work is that the magnetic signature of the Verwey transition is not unique: the $\mu$ vs. $T$ step at $T_{\rm V}$ may have opposite signs depending on the applied field and magnetic history. In finite fields of 10 mT and 0.18 or 0.25 T, $\mu$ increases upon warming through $T_{\rm V}$, whereas in nominally zero or very small fields, $B\leq 1$ mT, $\mu$ decreases. This field-dependent reversal of the magnetic step shows that the magnetization anomaly at $T_{\rm V}$ is not a simple marker of the electronic transition alone. Instead, it reflects the coupled response of magnetic domains, structural domains, anisotropy, dipolar interactions, and defect-induced pinning.

We also showed that magnetic-field-induced AS, namely the reorientation of the magnetic easy axis together with the monoclinic $c$ axis, has a particularly strong effect on $\mu(T)$ in stoichiometric magnetite. In this case, AS can make the transition appear broadened because the magnetic moment starts to change well below the true $T_{\rm V}$. In doped and nonstoichiometric samples this effect is smaller, but remains visible close to the transition. Thus, the apparent width and position of the magnetic anomaly can be shaped by magnetic and structural-domain processes rather than by chemical or electronic inhomogeneity alone.

These results clarify why different magnetic measurements can give different apparent signatures of the Verwey transition. They also show that the magnetization step, its sign, and its apparent temperature should be interpreted as diagnostics of the coupled magnetic--structural domain state, rather than as direct measures of the electronic transition alone. This distinction is important for understanding magnetite itself and for using magnetic measurements of the Verwey transition in studies of stoichiometry, defects, strain, and natural magnetite-bearing materials.

\begin{acknowledgments}
The authors acknowledge the Mineralogische Staatssammlung M{\"u}nchen, Museum ``Reich der Kristalle'' in Munich, Germany, for providing the natural magnetite sample. This work was supported by the National Science Centre, Poland, under OPUS Grant No. 2021/41/B/ST3/03454, and by the subsidy of the Ministry of Science and Higher Education of Poland. The research project was partly supported by the ``Excellence Initiative -- Research University'' program for the AGH University of Krakow, Project No. 6387. NB was supported by the Croatian Science Foundation under Project No. IP-2022-10-3382 and by the CeNIKS project, co-financed by the Croatian Government and the European Union through the European Regional Development Fund, Competitiveness and Cohesion Operational Program, Grant No. KK.01.1.1.02.0013. The work at TU Wien was supported by the Austrian Science Fund (FWF) [10.55776/F86; 10.55776/P35945].
\end{acknowledgments}

\bibliography{references}

\end{document}

\raggedbottom

\title{Supplemental Material: Interactions between structural and magnetic domains through the Verwey transition in stoichiometric and doped magnetite}

\author{Mateusz \'{S}l\k{e}zak}
\author{Zbigniew K\k{a}kol}
\author{Ryszard Zalecki}
\author{Andrzej Koz\l{}owski}
\affiliation{AGH University of Krakow, Faculty of Physics and Applied Computer Science, al. Mickiewicza 30, 30-059 Krakow, Poland}
\author{Naveen Kumar Chogondahalli Muniraju}
\affiliation{The Henryk Niewodniczanski Institute of Nuclear Physics, Polish Academy of Sciences, ul. Radzikowskiego 152, 31-342 Krakow, Poland}
\author{Neven Bari\v{s}i\'{c}}
\affiliation{Institute of Solid State Physics, TU Wien, 1040 Vienna, Austria}
\affiliation{Department of Physics, Faculty of Science, University of Zagreb, Bijenicka cesta 32, HR-10000 Zagreb, Croatia}
\author{Stuart Gilder}
\affiliation{Department of Earth and Environmental Sciences, Ludwig Maximilian University, Munich, Germany}
\author{Wojciech Tabi\'{s}}
\email{wtabis@agh.edu.pl}
\affiliation{AGH University of Krakow, Faculty of Physics and Applied Computer Science, al. Mickiewicza 30, 30-059 Krakow, Poland}

\maketitle

\renewcommand{\thefigure}{S\arabic{figure}}
\renewcommand{\thesection}{S\arabic{section}}

\section{Temperature dependence of magnetic anisotropy}

\begin{figure}[!htbp]
\centering
\includegraphics[width=0.95\textwidth]{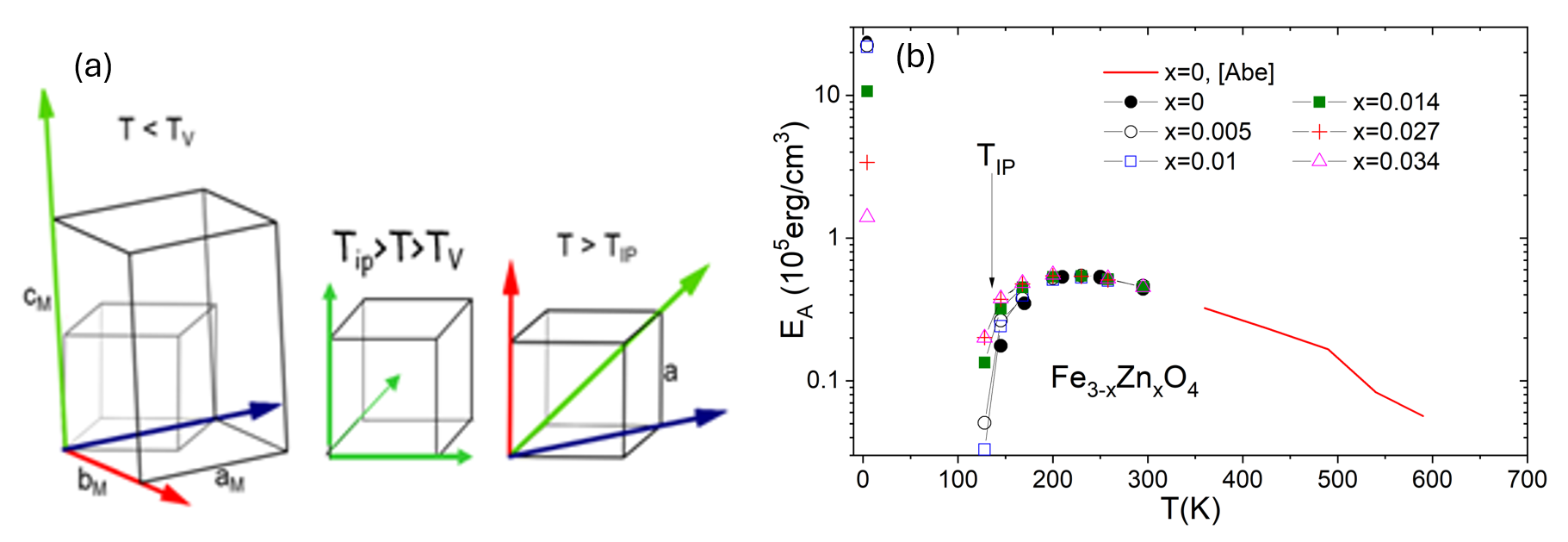}
\caption{\label{fig:S1}
Temperature dependence of magnetic anisotropy in magnetite. (a) Schematic temperature dependence of the main magnetic axes; the easy axes are shown in green. (b) Anisotropy energy of stoichiometric and Zn-doped magnetite, based on Refs.~\cite{abe1976,koenitzer1992}. The arrow marks the isotropy point $T_{\rm IP}$.}
\end{figure}

High-temperature magnetite has its easy, intermediate, and hard axes along the cubic $\langle 111\rangle$, $\langle 110\rangle$, and $\langle 001\rangle$ directions, respectively (Fig.~\ref{fig:S1}). This situation changes at the isotropy point $T_{\rm IP}\simeq 130$ K, where the anisotropy energy vanishes. Below this temperature, the magnetic moment points toward a $\langle 100\rangle$ direction. Below $T_{\rm V}$, the anisotropy energy in stoichiometric magnetite increases by roughly two orders of magnitude. A smaller, but still substantial, change occurs in doped samples, as shown for zinc ferrites in Fig.~\ref{fig:S1}. The easy magnetization axis is parallel to the monoclinic $c$ axis.

\section{Easy-axis switching and VSM geometry}

\begin{figure}[!htbp]
\centering
\includegraphics[width=0.85\textwidth]{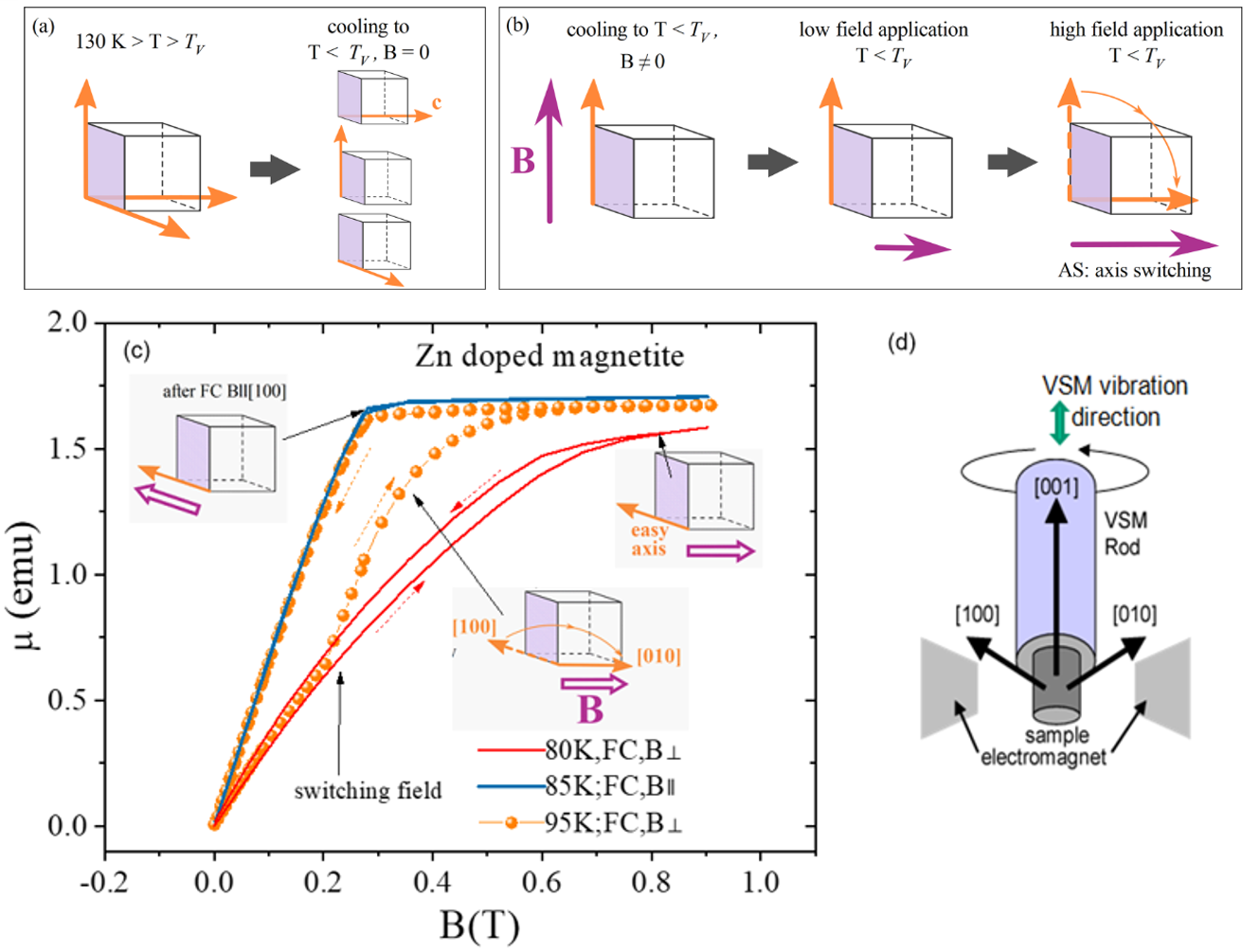}
\caption{\label{fig:S2}
Easy-axis switching and VSM geometry. (a) Effect of zero-field cooling on the magnetite structure; orange arrows denote possible monoclinic $c$ axes. (b) Effect of field cooling and subsequent field application: when the sample is heated in a field perpendicular to the initial $c$ axis, the $c$ axis may switch toward the field direction. (c) Axis switching observed in $\mu(B)$ for the Zn-doped sample 448\#1BW. When $B$ exceeds the strongly temperature-dependent switching field $B_{\rm sw}$, the $c$ axis switches, resulting in a sudden change in $\mu(B)$ and a response resembling that measured along the easy axis. (d) VSM geometry. The sample could be field-cooled along [100] and then measured either along [100] or along [010]. Figure adapted from Ref.~\cite{kolodziej2023}.}
\end{figure}

When magnetite is zero-field-cooled from the temperature range $130~{\rm K}>T>T_{\rm V}$, i.e., slightly below the isotropy point $T_{\rm IP}$, where the easy magnetic axes are along the cubic $\langle 100\rangle$ directions, several crystallographic domains form. In particular, monoclinic $c$ axes point along the former cubic $\langle 100\rangle$ directions, as indicated by the orange arrows in Fig.~\ref{fig:S2}a. In the case of field cooling along one of the cubic $\langle 100\rangle$ directions, the field direction becomes the monoclinic $c$ axis below $T_{\rm V}$ (Fig.~\ref{fig:S2}b). If a magnetic field is then applied perpendicular to this $c$ axis, the $c$ axis may switch toward the field direction. This phenomenon is referred to as axis switching (AS). The experimental realization of this effect for the Zn-doped sample 448\#1BW is shown in Fig.~\ref{fig:S2}c.

When a field is applied during heating to a ZFC sample, one $c$ axis is initially collinear with $B$, while $B$ is perpendicular to the remaining two $c$ axes. Once the field exceeds a certain value, or once the temperature is sufficiently close to $T_{\rm V}$ from below, the two perpendicular $c$ axes switch toward the field direction, resulting in one dominant $c$ axis and a change in magnetic moment.

The VSM sample geometry is shown in Fig.~\ref{fig:S2}d. The cylindrical sample was glued to the VSM rod with its cubic [001] direction vertical. The sample could then be field-cooled in 0.7 T along [100] and measured either along the easy axis [100] or along [010], which is magnetically and crystallographically unspecified immediately after FC but may become the $c$ axis if the magnetic field is sufficiently large. Alternatively, the sample could be zero-field-cooled, which results in several $c$-axis variants, as shown schematically in Fig.~\ref{fig:S2}a.

\section{Temperature and doping dependence of axis switching}

\begin{figure}[!htbp]
\centering
\includegraphics[width=0.75\textwidth]{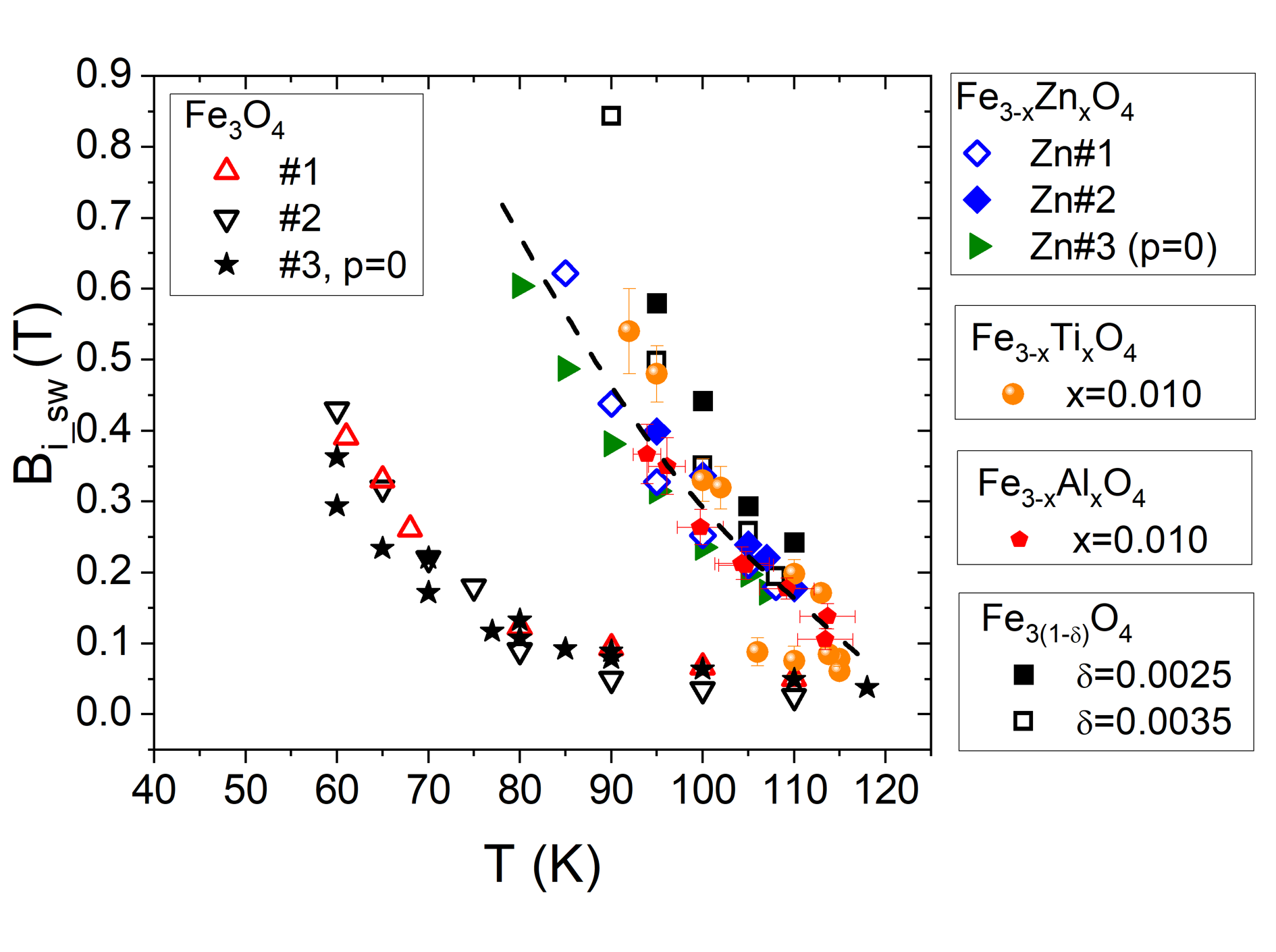}
\caption{\label{fig:S3}
Temperature dependence of the switching field for stoichiometric and doped magnetite crystals, based on Ref.~\cite{kolodziej2023}.}
\end{figure}

The switching field $B_{\rm sw}$, or the intrinsic switching field $B_{i,{\rm sw}}$ acting on the magnetic moments, depends strongly on nonstoichiometry and doping. Because most of the samples gathered in Fig.~\ref{fig:S3} have comparable defect content, a separate dependence of the switching field on defect content is not clearly resolved.

\section{Samples and their characterization}

The synthetic samples were at least 99.99\% pure single crystals grown at Purdue University by the skull-melter, crucibleless technique \cite{harrison1978}. This technique allows the oxygen partial pressure to be controlled during growth, thereby ensuring that the melt remains within the stability range of magnetite. After preparation, the crystals were subjected to subsolidus annealing under CO/CO$_2$ gas mixtures to establish the appropriate metal-to-oxygen ratio \cite{aragon1983,aragon1982,wang1987}.

\begin{figure}[!htbp]
\centering
\includegraphics[width=0.95\textwidth]{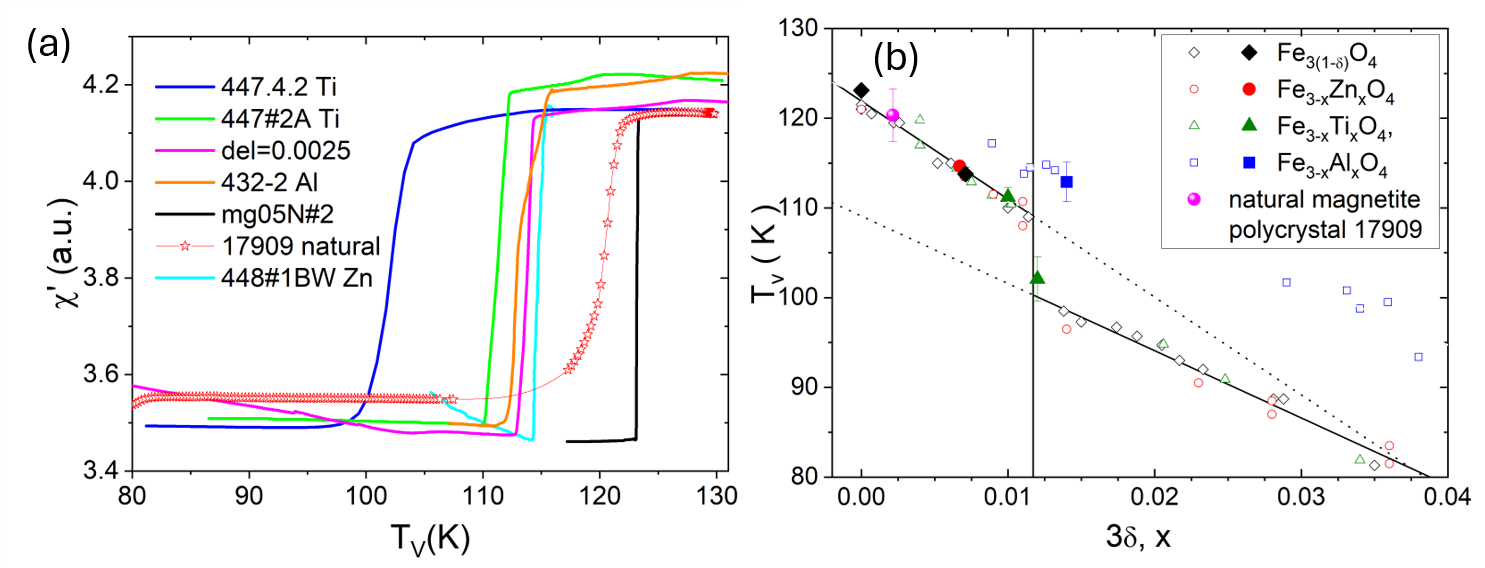}
\caption{\label{fig:S4}
Sample characterization. (a) Dispersive part of the ac magnetic susceptibility, $\chi'_{\rm AC}$, as a function of temperature for the studied magnetite samples. (b) Universal relation between $T_{\rm V}$ and the nonstoichiometry parameter $3\delta$ or doping level $x$; open symbols are literature data, and bold symbols mark the samples measured here. The $T_{\rm V}$ error bar corresponds to the transition width. The vertical line separates the discontinuous (left) and continuous Verwey-transition regimes. Literature curves and data are based on Refs.~\cite{kozlowski1996,kakol2015}.}
\end{figure}

The initial characterization of the samples was performed by ac magnetic susceptibility (Fig.~\ref{fig:S4}a). The dispersive part of the susceptibility, $\chi'_{\rm AC}$, was measured with a homemade susceptometer in the temperature range 77--140 K, with a peak ac-field amplitude of 0.05 mT and frequency 188.88 Hz. Except for one Ti-doped sample, all samples are only slightly doped and fall in the first-order VT regime; see Fig.~\ref{fig:S4}b, where the universal relation between $T_{\rm V}$ and either $x$ or $3\delta$, based on Refs.~\cite{kozlowski1996,kakol2015}, is shown.

\section{Detailed experimental protocol}

\begin{figure}[!htbp]
\centering
\includegraphics[width=0.95\textwidth]{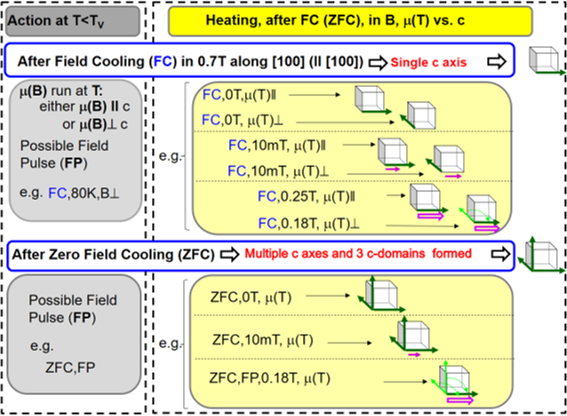}
\caption{\label{fig:S5}
Detailed experimental protocol used for the magnetization measurements. The scheme shows the FC and ZFC preparation branches, the optional low-temperature actions at 78 K, and the subsequent warming measurements of $\mu(T)$. Green arrows denote the monoclinic $c$ axis, which is also the magnetic easy axis below $T_{\rm V}$; purple arrows denote the applied field or field pulse, and open purple arrows indicate fields that can drive $c$-axis switching. The notation is described in the text.}
\end{figure}

The detailed protocol used for the magnetization measurements is summarized in Fig.~\ref{fig:S5}. The scheme separates the two preparation routes: field cooling, FC, and zero-field cooling, ZFC. It also shows the optional low-temperature actions performed at 78 K and the subsequent warming measurements of $\mu(T)$ in different magnetic fields.

In the FC protocol, the sample was cooled from above the Verwey transition to 78 K in a field of 0.7 T applied along the cubic [100] direction. This selects a single monoclinic $c$ axis below $T_{\rm V}$, which is also the magnetic easy axis. In Fig.~\ref{fig:S5}, this selected $c$ axis is indicated by the dark-green arrows. After FC, $\mu(B)$ could be measured either parallel to this axis, $B\parallel c$, or perpendicular to it, $B\perp c$. In some measurements, the sample was additionally subjected to a field pulse, FP, typically 0.9 T for 2 min at 78 K.

In the ZFC protocol, no field was applied during cooling. Consequently, several monoclinic $c$-axis variants can form below $T_{\rm V}$, as indicated schematically in the lower part of Fig.~\ref{fig:S5}. For ZFC samples, no $\mu(B)$ curves were recorded before warming. In some cases, however, an FP was applied at 78 K before the $\mu(T)$ measurement.

The right-hand part of Fig.~\ref{fig:S5} summarizes the warming measurements. After FC, $\mu(T)$ was measured in nominally zero field, 10 mT, or 0.18--0.25 T, with the measuring field either parallel or perpendicular to the $c$ axis selected during cooling. Thus, for FC data, the symbols $\parallel$ and $\perp$ refer to the orientation of the warming field relative to the FC-selected $c$ axis. After ZFC, no unique initial $c$ axis exists; therefore, the symbols $\parallel$ and $\perp$ are not used unless a field pulse was applied, in which case they refer to the FP direction. The purple arrows denote the applied magnetic field or field pulse, whereas the open purple arrows indicate the field direction that can drive $c$-axis switching during warming. The notation used throughout the manuscript follows this sequence. For example, FC, 80 K, $B\perp$ denotes a sample field-cooled to 78 K and then measured at 80 K with the magnetic field perpendicular to the $c$ axis imposed during FC.

\section{\texorpdfstring{$\mu$ vs. $B$}{mu vs B} for all samples}

\begin{figure}[!htbp]
\centering
\includegraphics[width=0.95\textwidth]{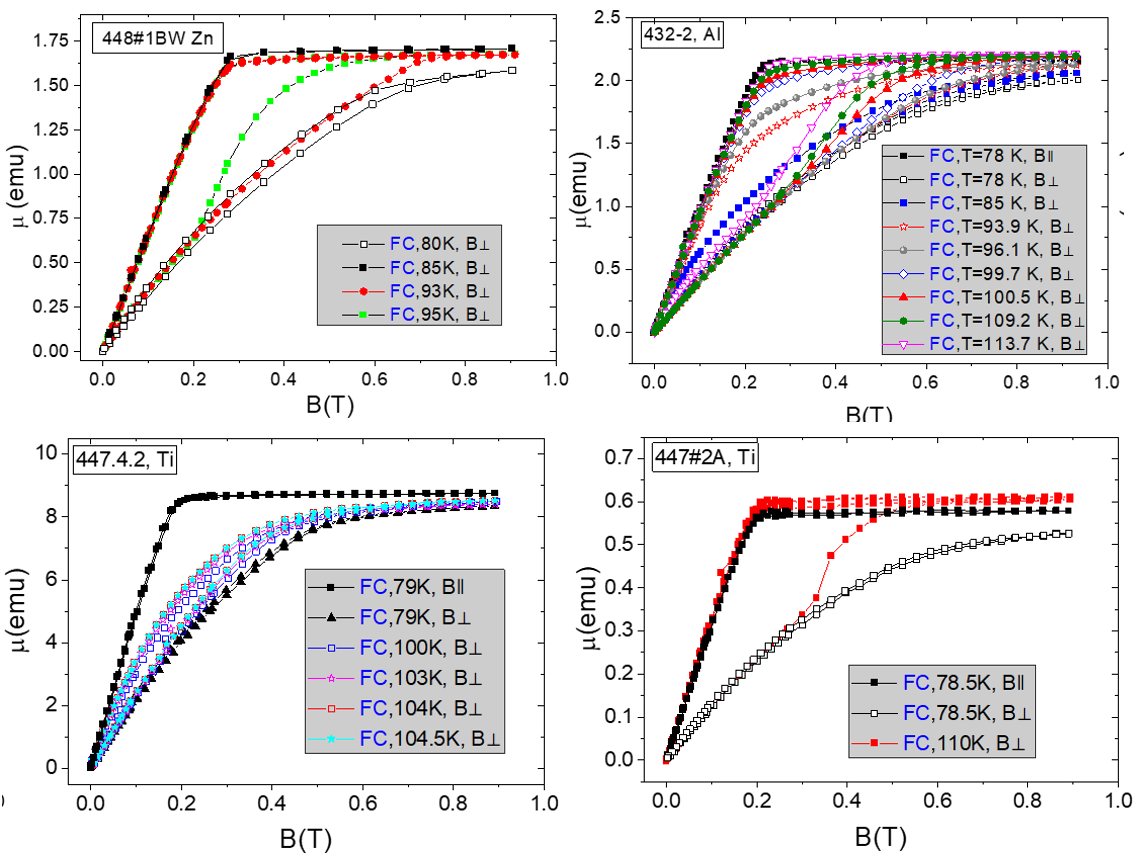}
\caption{\label{fig:S6}
$\mu$ vs. $B$ measured along the direction of the cooling field, i.e., [100], and along [010], which is magnetically and crystallographically unspecified immediately after FC. In almost all shown samples, and also for $\delta=0.0025$ and natural magnetite in the main text, $\mu(B)$ along [010] at 78 K is practically reversible, indicating almost no AS at this temperature. The exception is stoichiometric mg0, shown in the main text, where $\mu(B)$ already departs from the 10 K response at 78 K. Note also the different shape of $\mu(B)$ for 447.4.2, which exhibits a continuous VT, and for the natural sample shown in the main text; for both samples, no clear switching field could be defined.}
\end{figure}

The $\mu$ vs. $B$ results for the stoichiometric sample, the nonstoichiometric sample with $\delta=0.0025$, and the natural magnetite sample 17909 are presented in the main text. The corresponding data for the remaining samples are shown in Fig.~\ref{fig:S6}.

\section{\texorpdfstring{$\mu$ vs. temperature superimposed on $\mu(B)$}{mu vs temperature superimposed on mu(B)}}

\begin{figure}[!htbp]
\centering
\includegraphics[height=0.78\textheight,keepaspectratio]{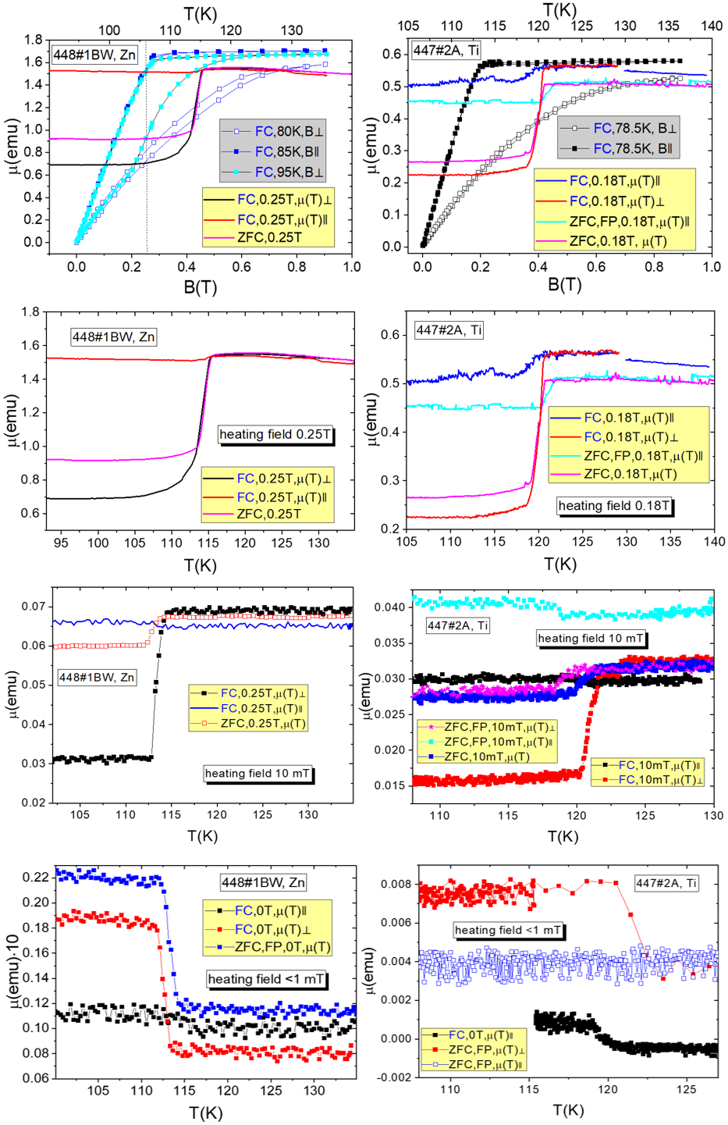}
\caption{\label{fig:S7}
Results of $\mu$ vs. $B$ and $\mu$ vs. $T$ measurements showing AS for the Zn-doped sample 448\#1BW and the Ti-doped sample 447\#2A. In the first row, the $\mu(B)$ results are superimposed on $\mu$ vs. $T$ in the transition region, showing the scale of $B$ in the experiment and the impact of AS on $\mu(T)$. The remaining rows show $\mu(T)$ for different fields. Note the difference in the $\mu$ vs. $T$ jump at $T_{\rm V}$ for $B<1$ mT, in contrast to $B=10$ mT and $B=0.25$ T.}
\end{figure}

\begin{figure}[!htbp]
\centering
\includegraphics[height=0.78\textheight,keepaspectratio]{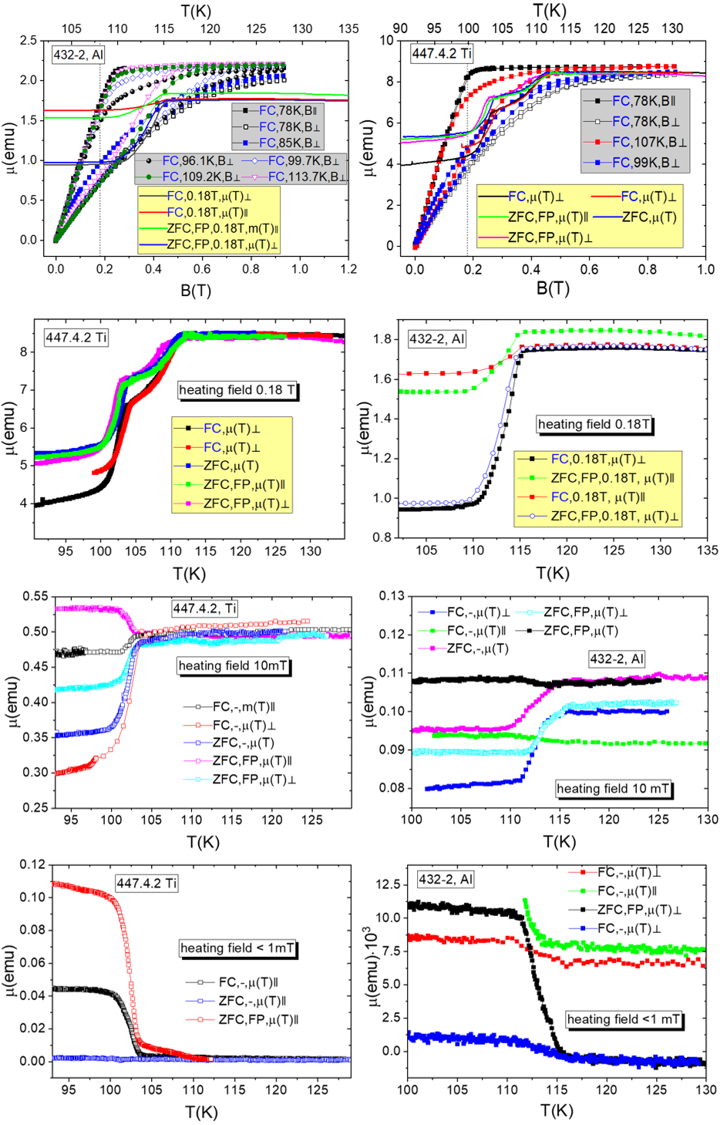}
\caption{\label{fig:S8}
Same as Fig.~\ref{fig:S7}, but for the Ti-doped sample 447.4.2 and the Al-doped sample 432-2.}
\end{figure}

The results for the stoichiometric sample, the nonstoichiometric sample with $\delta=0.0025$, and the natural magnetite sample 17909 are presented in the main text. The corresponding results for all other measured samples are shown in Figs.~\ref{fig:S7} and \ref{fig:S8}.

\section{Morphology of measured samples}

\begin{figure}[!htbp]
\centering
\includegraphics[width=0.95\textwidth]{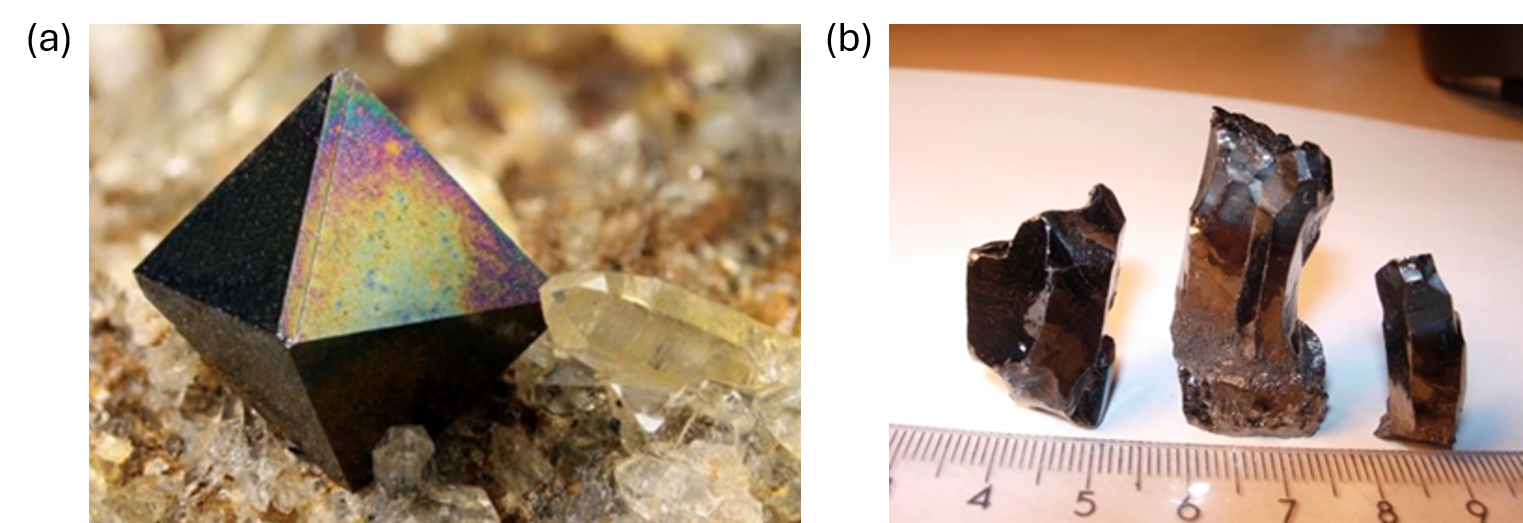}
\caption{\label{fig:S9}
Morphology of magnetite samples. (a) Natural magnetite single crystal. (b) Skull-melter-grown synthetic single crystals.}
\end{figure}

The synthetic magnetite crystals studied here were grown within one day, whereas natural magnetite can require millions or even billions of years to form. The most visible difference between these two types of material is their shape, shown in Fig.~\ref{fig:S9}. In contrast to the synthetic crystals measured here, impurities and structural complexity cannot be avoided in the natural crystal. This affects, for example, the field needed to switch the magnetic easy axis.

\section{Auxiliary ac susceptibility measurements of \texorpdfstring{$\delta=0.0025$}{delta=0.0025}}

\begin{figure}[!htbp]
\centering
\includegraphics[width=0.95\textwidth]{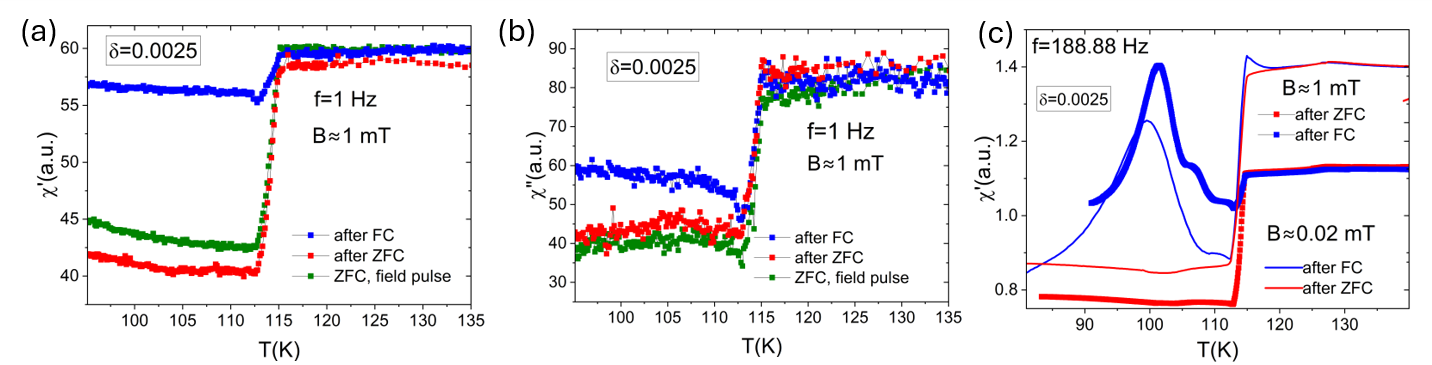}
\caption{\label{fig:S10}
Auxiliary $\chi_{\rm AC}$ measurements for the nonstoichiometric single crystal with $\delta=0.0025$. (a,b) Dispersive and absorptive parts of $\chi_{\rm AC}$ measured at an ac frequency of 1 Hz, representative also down to 0.2 Hz. (c) Dispersive part measured at 188.88 Hz. After FC, a large maximum appears approximately 15 K below the Verwey transition temperature, probably caused by point defects.}
\end{figure}

Alternating-current magnetic susceptibility $\chi_{\rm AC}$ was measured for the nonstoichiometric sample with $\delta=0.0025$ \cite{tabis} to check whether lowering the frequency down to 0.2 Hz, with cooling and heating treatments similar to those used in the VSM experiments, changes the susceptibility step from an up-step, as observed for $B>20$ mT, to a down-step, as observed for $B\leq 1$ mT. As shown in Fig.~\ref{fig:S10}, this did not occur. However, defects present in the sample due to nonstoichiometry changed the domain-wall motion, producing a large maximum approximately 15 K below the Verwey transition temperature when the sample was heated after 0.2 T field cooling.

\bibliographystyle{apsrev4-2}
\bibliography{references}